# Correlated states in twisted double bilayer graphene

Cheng Shen[1,2], Yanbang Chu[1,2], QuanSheng Wu[3,4], Na Li[1,2], Shuopei Wang[1,7], Yanchong Zhao[1,2], Jian Tang[1,2], Jieying Liu[1,2], Jinpeng Tian[1,2], Kenji Watanabe[5], Takashi Taniguchi[5], Rong Yang[1,6,7], Zi Yang Meng[1,7,8], Dongxia Shi[1,2,6], Oleg V. Yazyev[3,4] and Guangyu Zhang[1,2,6,7]*

[1]Beijing National Laboratory for Condensed Matter Physics and Institute of Physics, Chinese Academy of Sciences, Beijing 100190, China

[2]School of Physical Sciences, University of Chinese Academy of Sciences, Beijing 100190, China

[3]Institute of Physics, Ecole Polytechnique Fédérale de Lausanne (EPFL), CH-1015 Lausanne, Switzerland

[4]National Centre for Computational Design and Discovery of Novel Materials MARVEL, Ecole Polytechnique Fédérale de Lausanne (EPFL), CH-1015 Lausanne, Switzerland

[5]National Institute for Materials Science, 1-1 Namiki, Tsukuba 305-0044, Japan

[6]Beijing Key Laboratory for Nanomaterials and Nanodevices, Beijing 100190, China

[7]Songshan Lake Materials Laboratory, Dongguan, Guangdong 523808, China

[8]HKU-UCAS Joint Institute of Theoretical and Computational Physics，Department of Physics, The University of Hong Kong, Hong Kong, China

*Corresponding author. Email: gyzhang@iphy.ac.cn

**Electron-electron interactions play an important role in graphene and related systems and can induce exotic quantum states, especially in a stacked bilayer with a small twist angle.[1-7] For bilayer graphene where the two layers are twisted by a "magic angle", flat band and strong many-body effects lead to correlated insulating states and superconductivity.[4-7] In contrast to monolayer graphene, the band structure of untwisted bilayer graphene can be further tuned by a displacement field,[8,9,10] providing an extra degree of freedom to control the flat band that should appear when two bilayers are stacked on top of each other. Here, we report the discovery and characterization of such displacement-field tunable electronic phases in twisted double bilayer graphene. We observe insulating states at a half-filled conduction band in an intermediate range of displacement fields. Furthermore, the resistance gap in the correlated insulator increases with respect to the in-plane magnetic fields and we find that the g factor according to spin Zeeman effect is ~2, indicating spin polarization at half filling. These results establish the twisted double bilayer graphene as an easily tunable platform for exploring quantum many-body states.**

Twisted bilayer graphene (TBG) with a small twist angle $\theta$ exhibits a significantly reconstructed band structure.[1,2,3] In the vicinity of magic angle, i.e. $\theta \approx 1.1°$, strong interlayer hybridization leads to the formation of flat band with low energy and narrow bandwidth, and greatly enhances the electronic interaction effect[1,4-7] compared to graphene and bilayer graphene without twisting.[8] In TBG, correlated insulating states and unconventional superconductivity have been observed for a variety of partially filled bands[4-7] and novel phenomena such as ferromagnetism and quantum anomalous Hall effect,[7,11-13] topological phases[14,15] and features resembling in high-$T$c superconductor[5,16] have been intensively explored, consequently attracting considerable interest in both theoretical and experimental communities.[17-19] However, to observe these interesting correlation-induced phenomena, one has to control $\theta$ accurately enough in TBG, which puts strict constraints on device fabrication. Therefore, easier access to the flat band by alternative approaches is of high importance. Recently, correlated insulator and superconductivity have been observed in TBG with $\theta$>1.2° by exerting an external high pressure.[6] Besides, by applying vertical electrical fields, similar behavior can be achieved in ABC-stacked trilayer graphene on hexagonal boron nitride ($h$BN).[20-22] Unfortunately, exerting displacement field to modulate bandwidth of flat band in TBG has little consequence due to strong interlayer hybridization.[1,3,6]

Twisted double bilayer graphene (TDBG), on the other hand, is also likely to possess a flat band and display correlated phenomena. It is known that monolayer graphene possesses linearly dispersive energy bands, showing no dependence on the displacement fields. In contrast, Bernal(AB)-stacked bilayer graphene shows parabolic band dispersion and gap opening at charge neutral point (CNP) could be induced under displacement fields.[8-10] The gap reaches its minimum not rightly at, but close to the K point, displaying a sombrero dispersion of band structure,[8,10] which facilitates the formation and tunability of flat band in TDBG.

In this work, we report the successful control of the electronic phases in TDBG by vertical displacement fields. In a specific range of displacement fields, we observe correlated insulating states corresponding to the half-filled conduction band. Moreover, under parallel magnetic fields, we found that the correlated gap for the half-filled band increases, suggesting a spin-polarized ordering.

Fig. 1a shows the schematic structure of our dual-gate devices (see Methods for details about device fabrication). Under proper displacement fields, the reconstructed band structure of AB-stacked bilayer graphene, shows a more pronounced flatness in the top of valence band and bottom of conduction band, and thus facilitate the formation of flat band in TDBG (Fig. 1b). Fig. 1c shows a typical band structure resulted from tight-binding calculations based on *ab initio* parameters. We can clearly see the well-isolated flat conduction band under an intermediate displacement field.

We tested many dual-gated devices with $\theta$ varying from 0.98° to 1.33° to reveal the transport behavior in TDBG (Extended Data Fig. 1). All devices show single-particle gaps at superlattice density $n=\pm n_s$ on both of electron and hole branches (Fig. 1d and Extended Data Fig. 1) and displacement-field induced gaps at CNP due to the Bernal-stacked lattice structure in original bilayer graphene. Note that higher-order superlattice gaps at $n=\pm 3n_s$ can be observed in devices with a smaller $\theta$ around 1.0°. The twist angle can be extracted from its relevance with superlattice carrier density $n_s$ (See more details in Methods). The dual-gate structure makes it easier to independently tune carrier density as well as the displacement field across TDBG. Displacement field and carrier density are set by $D = \frac{1}{2}(D_b + D_t)$ and $n = (D_b - D_t)/e$, where $D_b = +\varepsilon_0 \varepsilon_{rb}(V_b - V_b^0)/d_b$ and $D_t = -\varepsilon_0 \varepsilon_{rt}(V_t - V_t^0)/d_t$. Here, $\varepsilon_{rb}$ and $\varepsilon_{rt}$ are relative dielectric constant for bottom dielectric layer with thickness $d_b$, and top with $d_t$, respectively. In our devices, dielectric layers are composed of $SiO_2$ and $h$BN(Fig. 1a), both of which share the same relative dielectric constant $\varepsilon_r \approx 3.9$ in our experiments. $\varepsilon_0$ is vacuum permittivity. $V_b^0$ and $V_t^0$ represent the offset gate voltage to reach the charge neutrality.

Under an intermediate $D$, we can see an obvious resistive state at $n=n_s/2$ where a moiré unit cell is filled with 2 electrons (Fig. 1d), i.e. the single-particle conduction band located between charge neutrality and conduction superlattice band is half-filled. The asymmetry between flat conduction and valence bands presumably arises from an intrinsic particle-hole asymmetry in original Bernal-stacked bilayer graphene[23]. Figure 1e shows definitely insulating behavior of the half-filling state below $T \approx 15K$ at $D/\varepsilon_0 = -0.4V/nm$ (marked in Extended Data Fig. 1) in a 1.33°-device. This characteristic temperature is higher than most of reported results in TBG.[4-7] Fitting through Arrhenius formula $R \sim \exp(\Delta/2kT)$, the gap of half-filling state in the 1.33°-device is estimated in Fig. 1f, showing strong dependence on displacement fields and a maximum value of ~3.2meV. Note that the fitted gap at half-filling also varies with twist angles. According to our results, regime of 1.2°-1.3° doubtlessly produces observable insulating behavior above liquid helium temperature.

The induced gap at half filling is widely observed within the regime of $|D|/\varepsilon_0 = 0.2V/nm \sim 0.6V/nm$ (Fig. 1e, Fig. 2a) and features non-monotonic change with respect to displacement fields. As $D$ is increased, single-particle gap at CNP ($\Delta_{CNP}$) increases monotonously while at $n=\pm n_s$ ($\Delta_{\pm ns}$) decreases because of the electrostatic potential difference (Extended Data Fig. 2 and Extended Data Fig. 3). The isolated flat band, flanked by these single-particle gaps at CNP and $\pm n_s$, is tightly related to their gap size. Our data confirm that the correlated gap at half filling develops only when $\Delta_{CNP}$ starts to be noticeable. At a higher $D$, superlattice gap $\Delta_{ns}$ closes, leading to flat conduction band overlapping with conduction superlattice bands and finally vanishing of the correlated gap at $n=n_s/2$ (see Methods and Extended Data Fig. 3). Such understanding also conforms to the observed absence of correlated gap at $n=-n_s/2$ since the small size of the single-particle gap $\Delta_{-ns}$ and its closing at a smaller $D$, prevent the isolation of valence flat band due to thermal activated inter-band hopping. The $D$ response of correlated gap still needs to be further revealed as higher $D$ may also broaden the width of flat band for effects like trigonal warping and over-bended sombrero band dispersion in the composed AB-stacked bilayer graphene. In Extended Data Fig. 3, our calculations truly show a wider bandwidth when $D$ is high enough.

In TBG, the flat band intrinsically derives from strong interlayer coupling. Under an external displacement field, electrons can be drawn towards positive gate, and repelled away from the negative one, thus inducing an asymmetric distribution of carriers between top and bottom graphene layers and destroying the expected correlated states, as observed in dual-gated TBG.[6] In TDBG, stronger screening and thus stronger layer asymmetry under displacement fields are expected. All our devices show half-filling correlated states constrained by unexpected resistive states, which are tuned by a single gate (Fig. 1d and Extended Data Fig. 1). The layer-polarized distribution of electrons abruptly eliminates the correlated gap at $D$ where single-particle gaps both at CNP and $n=n_s$ still persist (Fig. 1d and Fig. 1f). Such layer-polarization is also reflected by the fact that correlated state behaves less insulating at $D$ where electrons are polarized at disordered graphene layer (Extended Data Fig. 1).

The half-filling correlated insulating state is further signified by a sign change of transverse Hall resistance at $n=n_s/2$ (Fig. 2a and Fig. 2c). A new Fermi surface originates from the half-filling insulating state and acquires an effective Hall carrier density $n_H=n-n_s/2$ (Extended Data Fig. 4). When focusing on doped correlated states, sign change of the transverse magneto resistance flanks correlated states, showing electron-like quasiparticles from charge neutrality (correlated insulator) transformed into hole-like quasiparticles from correlated insulator (superlattice bands). While in Fig. 2a, the doped domes of correlated insulator are also constrained by these resistive states, displaying a behavior distinct from that tuned expectedly only by Fermi level and thus revealing impacts from layer polarization on doped domes.

Moreover, our data also reveal that valence bands are also subjected to layer polarization, shown by single-gated resistive states connecting gap edges of $\Delta_{CNP}$ and $\Delta_{-ns}$ (Fig. 1d).

The new Fermi surface originated from half-filling correlated states can also be identified by a new set of magneto oscillations which typically serves as a tool to obtain information about

Fermi surface and degeneracy associated with electronic degrees of freedom.[2-7] Fig. 2b explicitly shows one-sided Landau levels (LLs), emanating from half-filling correlated states in a 1.28°-device. A Landau level of possible $v=2$ occurs close to the half-filling correlated insulating state at higher perpendicular field $B_\perp$, tending to be signals of Hofstadter butterfly in our view. We present three reasons for this conjecture: 1) this LL doesn't hold down to a small $B_\perp$; 2) this LL also occurs with the same behavior when half-filling correlated insulating state doesn't exist at a larger $D$ (Extended Data Fig. 5); 3) a fractal LL of $v=4$ and Bloch band filling factor $s=1$ also develops at a same range of $B_\perp$ (Here the Hofstadter energy spectrum is described by Diophantine formula $n/n_0=v\varphi/\varphi_0+s$, where $n_0=1/A=n_s/4$, $\varphi=B_\perp A$ is the magnetic flux penetrating each moiré unit cell of area $A$, $\varphi_0=h/e$ is the non-superconducting quantum flux, and $s=\pm 4$ denotes that Landau fan results from superlattice band).[24-27] Two Landau levels of filling factor $v=3$ and $v=5$ fanning from half-filling correlated state are observed. Such complicated odd-number sequence of LLs from half-filling correlated insulator together with fractal LLs are checked carefully and repeated in the 1.33°-device (Extended Data Fig. 6) (we note that the fractal LL of $(v, s) = (4, 1)$ seems to be substituted by $(v, s) = (3, 1)$ in the 1.33°-device.

The LL filling factors of $v=3$ and $v=5$ is beyond our expectations considering that broken symmetry induces halved filling factors with even numbers for the LLs resulted from half-filling correlated insulating state in TBG.[5,6,7] However, under displacement fields, extra inversion-symmetry breaking is introduced and the Landau fan in Bernal-stacked bilayer graphene could be strongly affected as a consequence.[28] The LLs fanning from CNP in our TDBG devices, displays a fully lifted degeneracy in such a displacement field in Fig. 2b. While we also found other cases such as LL of $v=12$ at a lower $D$ (Extended Data Fig. 5) or LLs of $v=3$ and $v=5$ at $D/\varepsilon_0=-0.4$V/nm (Extended Data Fig. 6) dominating Landau fan diagram from CNP, which are also beyond the expected four-fold degeneracy, i.e. spin and valley without layer degeneracy due to the strong interlayer coupling. Likewise, the halved LL filling factor for half-filling correlated insulating state displays such $D$-dependent symmetry breaking.

The half-filling correlated state at $D/\varepsilon_0=-0.45$V/nm persists up to $B_\perp\approx 7$T according to the Hall coefficient $R_H$ behavior shown in Fig. 2c. It is visible in Fig. 2b and Extended Data Fig. 6 that the resistance of half-filling starts to increase and then decreases with respect to $B_\perp$. We further fitted the gap of half-filling correlated insulating state in the 1.33°-device (Extended Data Fig. 7). It exhibits remarkable enhancement as $B_\perp$ increases from 0T to 3T, then fades a lot at $B_\perp=6$T. Meanwhile, at $B_\perp=6$T, the correlated state at $3/4n_s$ develops, identified by resistive state, insulating behavior and potential sign change of $R_H$ (Fig. 2b, Fig. 2c, Extended Data Fig. 6 and Extended Data Fig. 7).

In order to isolate spin effects from the orbital motion, we thus exerted parallel magnetic fields and probed the field dependence of resistances. Fig. 3a shows vague resistive state but no insulating behavior at half-filling at $B_\parallel=0$T above $T=1.6$K in device $\theta=1.06°$ (graphite acts as back-gate). The parallel magnetic fields monotonously enhance resistance at $n=n_s/2$ and lead to an obvious insulating state at moderate $D$ and higher $B_\parallel$ (Fig. 3b and Fig. 3c). Since $B_\parallel$ only affects spin degree of freedom, the enhanced insulating behavior implies the insulating state at $n_s/2$ is likely to be spin-polarized. Arising from Zeeman effect, magnetic fields induce gap broadening $\Delta=g\mu_B B_\parallel$, between spin-up and spin-down electrons, where the $g$ factor for electrons in graphene is ~2 and $\mu_B$ is the Bohr magneton. From the Arrhenius formula of resistance, we obtain thermal activation gap as a function of $B_\parallel$, which shows nearly linear relationship (Fig. 3d). Thus, we deduce that the effective $g$ factor is about 2.12, coincident with the expectation. The possible totally spin-polarized ground state for half-filling correlated insulator in TDBG makes it reasonable that $B_\perp$ could also first enhance the gap at $n=n_s/2$ if spin effect surpasses orbital effect, in contrast to the case of TBG,[4,6] which shows decreased gap at half-filling both at $B_\parallel$ and $B_\perp$.

Note that we also observed quarter-filling correlated state formed and enhanced by $B_\parallel$ in the 1.31°-device, which is much more sensitive to displacement field (Extended Data Fig. 8). Compared with the 1.33°-device, the 1.31°-device shows much smaller fitted correlated gap due to stronger disorder, and finally leads to an underestimation of $g$ factor ≈0.825.

Fig. 4 shows $\rho_{xx}$–$T$ behavior at various carrier densities with $D/\varepsilon_0$ = -0.4V/nm in the 1.33°-device, where $\rho_{xx}$ is the four-probe resistivity. In proximity to the half-filling state, the resistivity behaves abrupt dropping-down onset at $T$≈12K and linearly reduced at higher temperatures, well distinct with the case far away from $n=n_s/2$ (Fig. 4b). This $\rho_{xx}$–$T$ behavior observed in TBG and ABC-stacked trilayer graphene is a signature of superconductivity.[5-7,21] The linear relationship between $\rho_{xx}$ and $T$ observed in TBG is likely to support electron-phonon scattering or strange metal behavior but still under debate.[16,29] In our TDBG 1.33°-device, linearity coefficient of $d\rho_{xx}/dT$ for $n=1.757\times10^{12}$cm$^{-2}$ and $n=2.3\times10^{12}$cm$^{-2}$ is about 90 Ω/K and small saturated resistivity, e.g. ~400 Ω, persists at the lowest temperature (≈1.5 K). Multi-probe measurements suggest that this device is actually composed by a majority of $\theta$≈1.33° together with a minority of $\theta$≈1.1° in series (Extended Data Fig. 9). The twist-angle inhomogeneity is likely the reason why the resistance cannot reach to zero in this device. We did preliminarily observe zero resistance in the 1.28°-device (Extended Data Fig.10), but more measurements are required to verify the presence of superconductivity at doped correlated insulating state in TDBG.

Our work demonstrates electrically tunable correlated states in TDBG. The hypothesized spin-polarized ground state at half-filling, different from that in TBG, reveals an important role played by layer numbers of the original constituent 2D materials in twist system. Novel electronic states like electrically tunable ferromagnetic Mott insulator, [30] Chern bands[15] and spin-triplet topological superconductivity, [31] are potentially present in TDBG and calling for further theoretical and experimental efforts to reveal the underlying mechanism.

*Note added*: during the revising of our manuscript, we have been aware of two related studies of ref. 32 and ref. 33.

## Methods

### Device fabrication

Twisted double bilayer graphene (TDBG) devices were fabricated following a typical "tear and stack" technique.[34] Raw materials of bilayer graphene, hexagonal boron nitride (20-35nm in thickness) and graphite flakes were first exfoliated on $SiO_2$ (300nm thick), then annealed in $Ar/H_2$ mixture at temperature up to 450°C for cleanness. Usually, moderate $H_2$ plasma etching was also applied to fully get rid of contaminations coming from the exfoliation process. We used Poly (Bisphenol A carbonate) (PC) supported by Polydimethylsiloxane (PDMS) on glass slide to pick up hexagonal boron nitride (hBN) firstly and then tear and pick up bilayer graphene. The home-made micro-position stage can control the rotation angle at 0.1° error range. We performed no annealing for TDBG as it tends to relax to nearly twist angle $\theta=0°$ once temperature is high. The fabrication of metal top gate and electrodes follows a standard e-beam lithography and e-beam metal evaporation. Devices were designed as Hall bar structure and shaped by traditional reactive ion etching with $CHF_3$ and $O_2$ mixture gas. Here, the metal top gate also acts as a mask for etching to ensure the channel is fully gated. Finally, all the bars were contacted through one-dimensional edge contact with Cr/Au electrodes.[35]

### Transport measurements

Transport measurements were performed in cryostat with a base temperature of 1.5K. We applied standard lock-in techniques with 31Hz excitation frequency and 1nA alternating excitation current or less than 200μV alternating excitation bias voltage which was achieved by a 1/1000 voltage divider to measure the resistance. All the transport data were acquired by four-terminal measurements.

### Twist angle extraction

A large range of twist angle was achieved in our TDBG devices. Note that the encapsulated structure prevents traditional probe-characterizing methods to detect twist angle $\theta$. We have to extract $\theta$ from transport data acquired at cryogenic temperature. Before loaded in cryostat, devices are first picked out at room temperature. The flat band present in magic-angle twisted graphene superlattice, greatly lowers down the carrier mobility, thus causes charge neutral point (CNP) unrecognized in $R$-$V_g$ or $G$-$V_g$ curves, i.e. "U" rather than "V" shape appears at room temperature.

For a small $\theta$, the calculation of $\theta$ follows the formula:

$$A = \frac{4}{n_s} = \frac{\sqrt{3}}{2}\lambda^2 \approx \frac{\sqrt{3}a^2}{2\theta^2},$$

where $A$ is moiré unit cell area, $\lambda$ is moiré wave length, $a$=0.246nm is the lattice constant of graphene and $n_s$ is the carrier density at which the flat conduction band is fulfilled. Generally, in TDBG, we still persist the view that four electrons or holes in a moiré unit cell fulfill the first conduction or valence band for four-fold spin and valley degeneracy, yielding single-particle superlattice gaps. In most of cases, estimating $\theta$ through $n_s$ shows some degree of errors, since superlattice gaps are usually present over a range of $n_s$. Instead, we tend to determine it through the carrier density $n_s/2$ for half-filling correlated insulating state. In devices with well-developed quantum oscillations, an additional alternative is to get moiré unit cell area directly through Landau level crossing at magnetic flux $\phi=B_\perp A=\phi_0/q$. where $B_\perp$ is perpendicular magnetic field, $q$ is an integer and $\phi_0=h/e$ is the non-superconductivity magnetic flux.

### *Insulating states in devices with various twist angles*

The half-filling correlated state has been observed in our devices with $\theta$ varied from 0.98° to 1.33° (Extended Data Fig. 1). Note that in devices with $\theta=0.98°$ and $\theta=1.06°$, vague resistance peaks, yet without insulating behavior, develop at half filling above $T=1.6$K (also see Fig. 3a). While in strong in-plane magnetic fields, these two devices can develop insulating behavior at half-filling above $T=1.6$K (Fig. 3c). For 1.26°, 1.28° and 1.33° devices, altering the direction of displacement fields, half-filling correlated state shows different resistance, which is considered as a sign of layer asymmetry in carrier distribution. Interestingly, the superlattice gap $\Delta_{-ns}$ at $n=-n_s$ exhibits "reentrant" behavior as displacement field $D$ is tuned over a large range, i.e. the strengthening of $D$ first switches off this gap and then switches on it. Moreover, single-particle gaps occur at $\pm 3n_s$, where 12 electrons or holes fill a moiré unit cell. Such phenomenon quite resembles the case of high-order superlattice gaps in graphene/hBN superlattice,[36] and is actually consistent with our calculation results (see Extended Data Fig. 3).

### *Single-particle gaps tuned by displacement fields in 0.98°-device*

In the 0.98°-device, we measured single-particle gaps with respect to displacement fields $D$ at $n=0$ ($\Delta_{\text{CNP}}$), $n=\pm n_s$ ($\Delta_{ns}$ and $\Delta_{-ns}$) and $n=\pm 3n_s$ ($\Delta_{3ns}$ and $\Delta_{-3ns}$). All single-particle superlattice gaps reach their highest values at $D/\varepsilon_0=0$ and then decrease in displacement fields. In the regime of $|D|/\varepsilon_0=0.3$V/nm~0.5V/nm, which is rightly the interval where the half-filling insulating states can be observed in devices with $\theta=1.2°$~1.3°, $\Delta_{ns}$ and $\Delta_{-ns}$ drop down to zero or nearly zero, preventing the isolation of the first conduction and valence bands. The coexistence of $\Delta_{ns}$ and $\Delta_{\text{CNP}}$ just happens at a narrow range of $D$, and within this range, both $\Delta_{ns}$ and $\Delta_{\text{CNP}}$ are with low value. Small neighbor single-particle gaps for such an angle $\theta=0.98°$, make thermal-activated interband hopping considerable at our base temperature, and finally contribute to the absence of insulating behavior for the half-filled conduction band. These data acquired in the 0.98°-device, reveal the importance of neighbor single-particle gaps for correlated insulator. In addition, $\Delta_{-ns}$ appears again at $|D|/\varepsilon_0>0.5$V/nm, providing the possibility of well-isolated flat valence band formation at high $D$.

### *Band structure calculations*

We use the extended tight-binding model to calculate the band structures of twisted double bilayer graphene. Extended Data Fig. 3 shows the evolution of band structure as the displacement field is increased. Besides the isolated flat conduction band and evolution of single-particle gaps at CNP and $\pm n_s$ which are consistent with our experimental observations, the calculated results also provide information about the relationship between the bandwidth of isolated conduction band and $D$. When $D$ is high enough, the calculated bandwidth becomes so large (>20meV) that it is comparable to the on-site Coulomb repulsion energy and thus likely leads to the absence of correlated insulating state.

In our calculations, atomic positions of TDBG are full relaxed with the classical force-field approach implemented in the LAMMPS package.[37] The second generation REBO potential[38] and the Kolmogorov-Crespi (KC) potential[39] were used to describe the intra-layer and inter-layer interactions. Only the $p_z$ orbital of carbon atom is considered into our tight-binding model $H = \sum_i \epsilon_i a_i^\dagger a_i + \sum_{i \neq j} V_{ij} a_i^\dagger a_j$, where $a_i^\dagger, a_i$ are the creation and annihilation operators. $V_{ij} = V_{pp\pi} \sin^2\theta + V_{pp\sigma} \cos^2\theta$, where $\theta$ is the angle between the orbital axes and $\boldsymbol{R}_{ij} = \boldsymbol{R}_i - \boldsymbol{R}_j$ connects the two orbital centers. $\theta = \pi/2$ ($\theta = 0$) corresponones to the pair of atoms in the same layer (the pair of atoms on top of each other). The Slater-Koster[40] parameters $V_{pp\pi}$ and $V_{pp\sigma}$ depend on the distance $r$ between two orbitals as $V_{pp\sigma}(r) = V_{pp\sigma}^0 e^{q_\sigma(1-\frac{r}{a_\sigma})} F_c(r)$ and $V_{pp\pi}(r) = V_{pp\pi}^0 e^{q_\pi(1-\frac{r}{a_\pi})} F_c(r)$.[41] In our calculation, we use $V_{pp\pi}^0 = -2.81eV$, $V_{pp\sigma}^0 = 0.48eV$, $a_\sigma = 3.349Å$, $q_\sigma = 7.428$, $a_\pi = 1.418Å$, $q_\pi = 3.1451$, $F_c(r) = (1 + e^{(r-r_c)/l_c})^{-1}$, $l_c = 0.265Å$, $r_c = 6.165Å$.[42] The electric field is added through the onsite energy $\epsilon_i = Ez_i$, where $E$ is the strength of electric field, $z_i$ is the atomic z-axis coordinate. The band structures are obtained with the WannierTools open-source software package.[43]

*Extended data for quantum oscillations*

We repeatedly observed the magnetoresistance (SdH) oscillations near half-filling in the 1.33°-device. Extended Data Fig. 6 shows the same Landau levels (LLs) of filling factor $v=3$ and $v=5$ originated from half-filling correlated state. The differences between oscillation features shown in Fig. 2b for the 1.28°-device and Extended Data Fig. 6 for the 1.33°-device, are Landau level structures around CNP and Hofstadter butterfly patterns. In Extended Data Fig. 6, LLs near CNP show dominated filling factor $v=3$ and $v=5$ and weaker $v=4$, possibly because of a weaker displacement field compared with that in Fig. 2b. The Hofstadter butterfly in Extended Data Fig. 6, is established by a series of fractal LLs of $(v, s) = (3, 1), (2, 2), (1, 3)$ and $(2, 3)$ (Here, $v$ and $s$ are with the same meanings as described in the main text). Yet, the fractal LL $(v, s) = (3, 1)$ in Extended Data Fig. 6 is replaced by $(v, s) = (4, 1)$ in Fig. 2b.

The Landau level structure is affected by displacement field $D$. In a weaker field $D$, LLs from CNP show four-fold degeneracy on hole branch and dominated $v=12$ on electron branch (Extended Data Fig. 5c). While in a larger field $D$, degeneracy is fully lifted. For Hofstadter butterfly, likewise, a larger $D$ induces more visible fractal LLs displayed in Extended Data Fig. 5e, compared with the case in a weaker field that only LL crossing and unrecognized fractal LL are developed.

*Quarter-filling correlated insulating state induced by parallel magnetic fields*

The in-plane magnetic field strengthens half-filling insulator, also induces quarter-filling correlated insulating state as shown in Extended Data Fig. 8 in 1.31°-device. Stronger disorders in 1.31°-device lead to an underestimation of the thermal activation gap and finally a fitted $g$ factor less than 2. Likewise, the disability to observe $B_\parallel$-induced spin-polarized 3/4-filling correlated state could be also attributed to the bad quality of 1.31°-device.

40     Slater, J. C. & Koster, G. F. Simplified LCAO method for the periodic potential problem. *Phys. Rev.* **94**, 1498(1954).

41     Trambly de Laissardière, G., Mayou, D. and Magaud, L. Numerical studies of confined states in rotated bilayers of graphene *Phys. Rev. B* **86**, 125413(2012).

42     Haddadi, F., Wu, Q., Kruchkov, A. J., Yazyev, O. V., Moiré flat bands in twisted double bilayer graphene. Preprint at https://arxiv.org/abs/1906.00623(2019).

43     Wu, Q., Zhang, S., Song, H.-F., Troyer, M. & Soluyanov, A. A. WannierTools: An open-source software package for novel topological materials. *Comput. Phys. Commun.* **224**, 405-416(2018).

**Acknowledgements**

We appreciate the helpful discussion with Guorui Chen in UC Berkeley, Gaopei Pan and Shiliang Li at Institute of Physics, Chinese Academy of Scineces (IOP, CAS) and also the help of transport measurements from Fan Gao and Yongqin Li at IOP, CAS. G.Z. thanks the finical supports from NSFC under the grant No. 11834017 and 61888102, the Strategic Priority Research Program of CAS under the grant No. XDB30000000, the Key Research Program of Frontier Sciences of CAS under the grant No. QYZDB-SSW-SLH004, and the National Key R&D program under grant No. 2016YFA0300904. Q.W. and O.V.Y. acknowledge support from NCCR MARVEL. Z.Y.M. acknowledges supports from the National Key R&D Program (2016YFA0300502), the Strategic Priority Research Program of CAS (XDB28000000), the NSFC (11574359) and the Research Grants Council of Hong Kong Special Administrative Region of China (17303019). K.W. and T.T. acknowledge supports from the Elemental Strategy Initiative conducted by the MEXT, Japan, A3 Foresight by JSPS and the CREST (JPMJCR15F3), JST. Numerical calculations were performed at the Swiss National Supercomputing Center (CSCS) under Project No. s832, the facilities of Scientific IT and Application Support Center of EPFL, the Center for Quantum Simulation Sciences in the Institute of Physics, Chinese Academy of Sciences, the Computational Initiative at the Faculty of Science at the University of Hong Kong and the Platform for Data-Driven Computational Materials Discovery at the Songshan Lake Materials Laboratory, Guangdong, China.

**Data Availability**

The data represented in Fig. 1c-f, Fig. 2a-c, Fig .3a-d and Fig.4 are provided with the paper as Source Data. All other data that support the plots within this paper and other findings of this study are available from the corresponding author upon reasonable request.

**Author contributions**

G.Z. supervised this work. C.S. conceived of this project. C.S. and Y.C. fabricated the devices. C.S. performed the transport measurements and data analysis. Q.W. and O.V.Y. provided numerical calculations. K.W. and T.T. provided hexagonal boron nitride crystals. C.S., Z.Y.M. and G.Z. wrote the paper. All the other authors are involved in the discussion on this work.

**Competing interests**

The authors declare no competing interests.

# Figures and Captions

**Fig. 1| Correlated insulating states at half-filling tuned by displacement fields. a**, Schematic of moiré unit cell in twisted double bilayer graphene (TDBG) (left) and device structure (right). **b**, Pedagogical illustration of electrical tuning mechanism. Two parabolic-dispersive bands from original Bernal-stacked bilayer graphene layers are displaced by the twist angle $\theta$ (left, without displacement field $D$); and "Mexican-hat"-shaped bands in each bilayer graphene are induced by nonzero $D$ (right). The middle figure displays the cross-section configuration of AB-AB stacked TDBG. **c**, Calculated band structure of TDBG with $\theta=1.3°$ in vertical electrical field $|E|=50$mV/nm, which corresponds to $|D|/\varepsilon_0\approx0.2$V/nm by adopting a dielectric constant of $h$BN without considering the screening effect in TDBG. The horizontal color bars indicate various single-particle band gaps. The right figure plots density of state (DOS) at different energy. Here, $A$ is the area of moiré unit cell. **d**, Color plot of four-probe resistance versus metal top-gate and Si back-gate voltages in the 1.28°-device at $T=4$K. **e**, Temperature-dependent four-probe resistivity $\rho_{xx}$ versus carrier density $n$ in the 1.33°-device. The temperature $T$ varies from 1.7K (dark blue curve) to 14.6K (dark red curve). **f**, Thermal-activation gap at half-filling in the 1.33°-device as a function of displacement field $D$. The gap is fitted with Arrhenius formula $R\sim\exp(\Delta/2kT)$. The dash line denotes nearly linear enhancement of the half-filling correlated gap with respect to $D$. Error bars are estimated from the uncertainty in the range of simply activated regime.

**Fig. 2| Hall resistances and Quantum oscillations. a,** Transverse Hall resistance $R_{xy}$ as a function of carrier density $n$ and displacement field $D$ in the 1.31°-device with perpendicular magnetic field $B_\perp=2.25$T applied. In this color plot, red color represents positive $R_{xy}$ and indicates hole-type Hall carriers, conversely for blue color. **b, d**, Magnetoresistance (SdH) oscillations in the 1.28°-device at $T=3$K and $B_\perp$ varies from 0.9T to 9T. In our measurements, $V_{tg}$ is fixed at 8V and $V_{bg}$ is swept, producing $D/\varepsilon_0=-0.45$V/nm at half-filling, $D/\varepsilon_0=-0.35$V/nm at 3/4 filling and $D/\varepsilon_0=-0.65$V/nm at charge neutrality point (CNP). The sweeping trace is marked in Extended Data Fig. 5. $D$-fixed sweeping is given in the 1.33°-device, producing the same Landau levels (LLs) of $\nu=3$ and $\nu=5$ from half-filling at $D/\varepsilon_0=-0.4$V/nm (Extended Data Fig. 6). **c**, Hall coefficient $R_H$ ($R_H = \frac{(R_{xy}(B_\perp)-R_{xy}(B_\perp=0T))}{B_\perp}$) versus carrier density $n$ in various perpendicular magnetic fields in the 1.28°-device. A sign change at half-filling persists to $B_\perp=6$T and disappears at $B_\perp=8$T.

**Fig. 3| Parallel-magnetic-field response of half-filling correlated insulator. a, b**, Resistance color plot as a function of top and bottom metal gate voltage for the 1.06°-device in parallel magnetic fields $B_\parallel=0$T (**a**) and $B_\parallel=9$T (**b**). **a** and **b** share the same color scale bar. The dark dash lines in **a** and **b** represent resistive states at CNP and $n_s$. When $B_\parallel=9$T is applied, resistive state at $n_s/2$ appears. Top and back gate voltage sweeping along the green dot line in **b**, keeps $D/\varepsilon_0=-0.306$V/nm and linearly changes $n$. **c**, Resistance at $D/\varepsilon_0=-0.306$V/nm enhanced by parallel magnetic fields. An insulating behavior is shown in the top figure when $B_\parallel=9$T and $T<7.5$K but a metal behavior in the bottom figure instead when $B_\parallel=0$T and $T>1.5$K. **d**, Thermal activation gap $\Delta$ of $n_s/2$ insulating state as a function of $B_\parallel$. A linear fitting displayed by orange dash line gives effective $g$ factor ~2.12 according to the Zeeman effect $\Delta=g\mu_B B_\parallel$. The gap is extracted from fitting the data with $R\sim\exp(\Delta/2kT)$ in the bottom inset figure. The top inset figure illustrates that single-particle flat band is split into upper and lower spin-polarized many-body bands by e-e interactions. In parallel magnetic fields, each band contributes to gap broadening by $g\mu_B B_\parallel/2$. Error bars are estimated from the uncertainty in the range of simply activated regime.

**Fig. 4| $\rho_{xx}$–$T$ behaviors in the 1.33°-device. a**, Four-probe resistance as a function of temperature $T$ and carrier density $n$ in the 1.33°-device when $D/\varepsilon_0=-0.4$V/nm. The correlated insulating states (CS) is flanked by two low-resistivity domes. **b**, $\rho_{xx}$–$T$ curves at various $n$ plotted with different colors corresponding to the labels in **a**. A dash line linearly fitting $\rho_{xx}$–$T$ at $n=1.757\times10^{12}$cm$^{-2}$, gives the linearity coefficient $d\rho_{xx}/dT=90\Omega$/K.

**Fig.1**

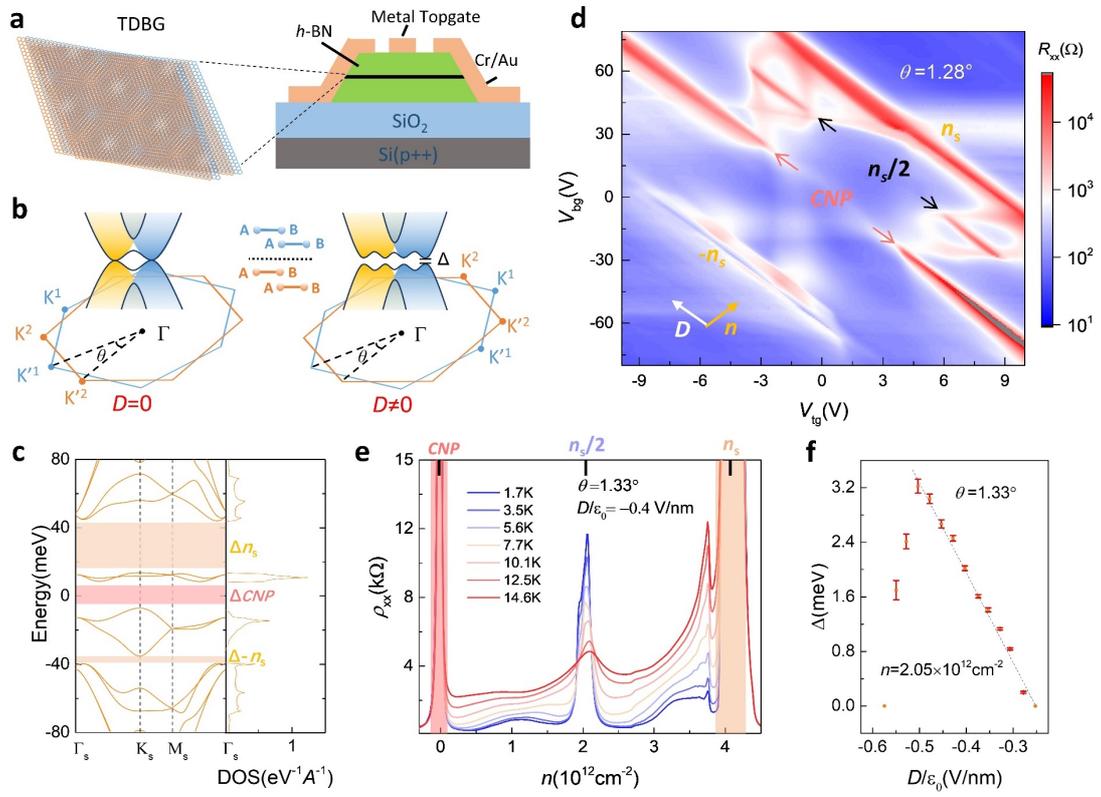

**Fig.2**

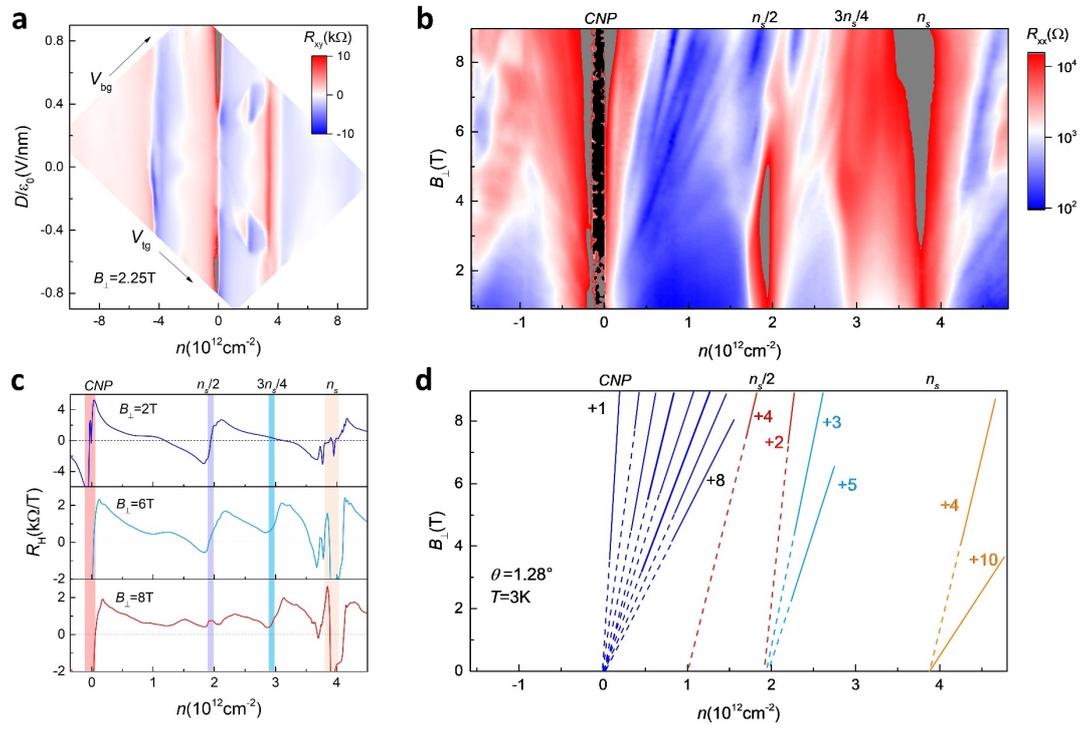

**Fig.3**

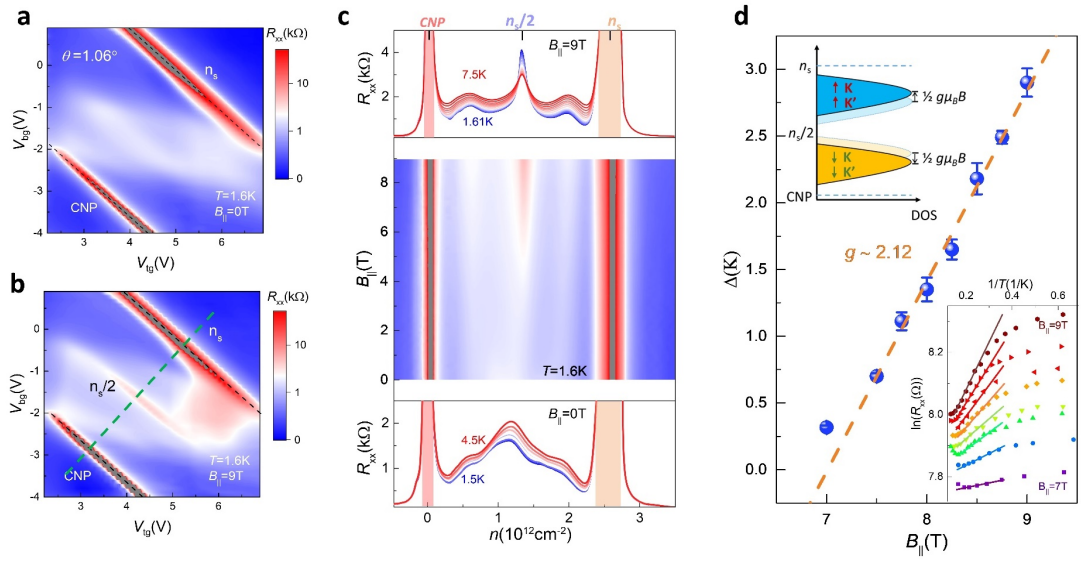

**Fig.4**

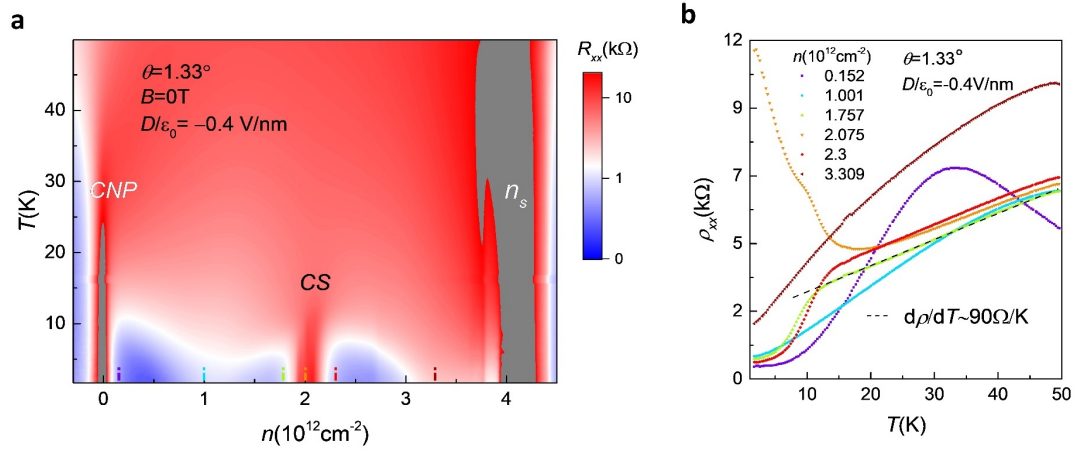

# Extended Data

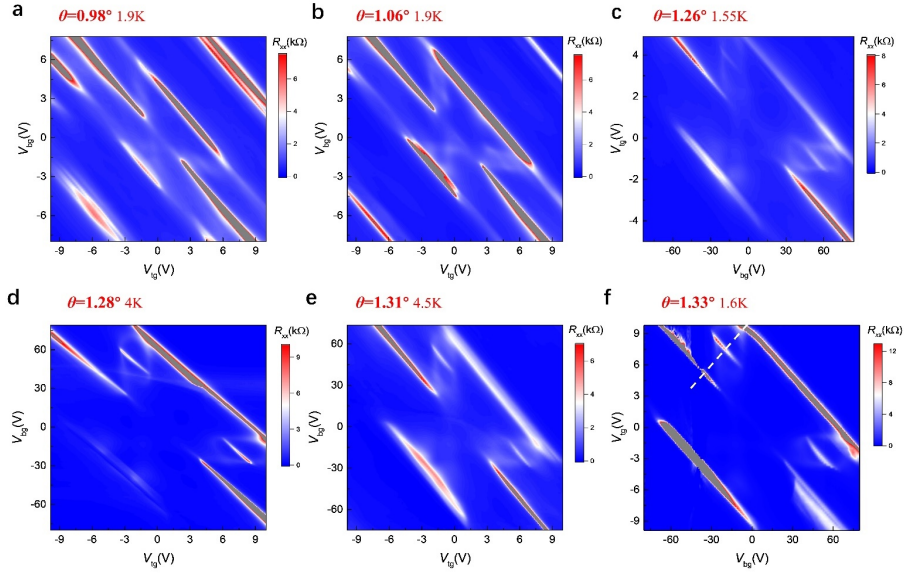

**Extended Data Fig. 1| Resistance mapping plots of all devices.** The resistance is plotted as a function of top and back gate voltages. Devices in our studies varies from 0.98° to 1.33°. The white dash line in **f** marks $D/\varepsilon_0$ = -0.4V/nm.

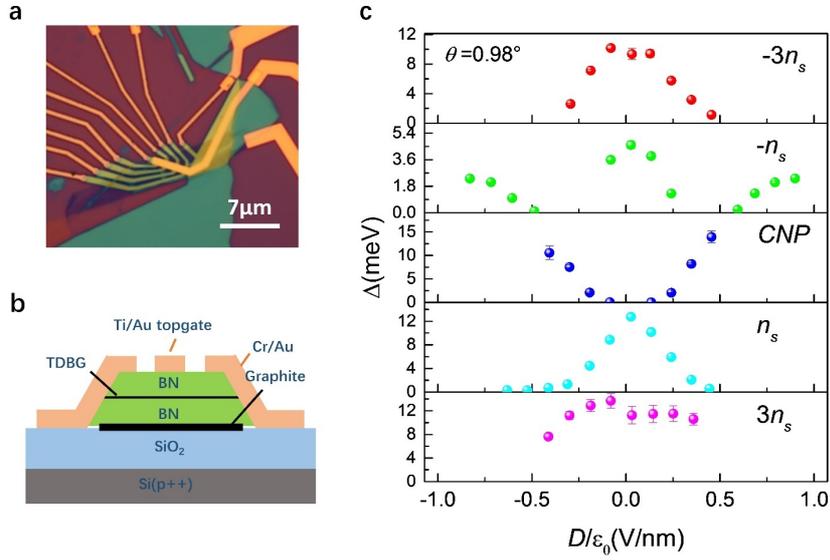

**Extended Data Fig. 2| Single-particle band gaps in the 0.98°-device. a**, Optical microscope image of the 0.98°-device. **b**, Schematic of device structure. **c**, Single-particle gaps at $n=0$, $n=\pm n_s$ and $n=\pm 3n_s$ with respect to displacement field. The thermal-activation gaps are fitted with Arrhenius formula $R\sim\exp(\Delta/2kT)$. Error bars are estimated from the uncertainty in the range of simply activated regime.

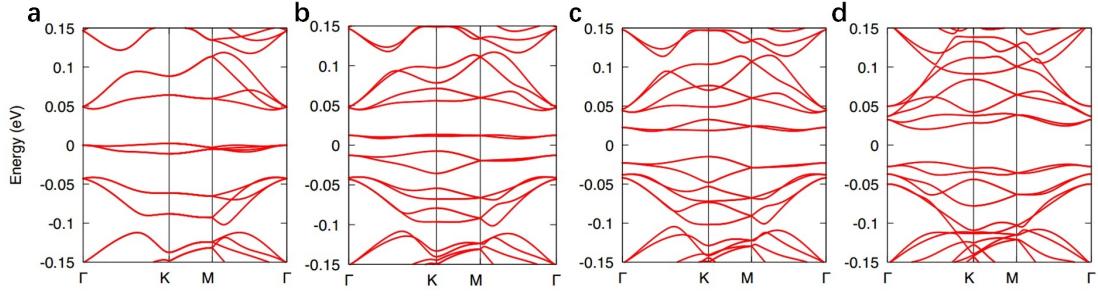

**Extended Data Fig. 3| Calculated band structures of 1.3° twisted double bilayer graphene in various displacement fields.** The electric fields (and corresponding displacement fields calculated with the relative dielectric constant of $h$BN) in **a**, **b**, **c** and **d** are 0, $|E|$=50mV/nm ($|D|/\varepsilon_0 \approx$ 0.2V/nm), $|E|$=90mV/nm ($|D|/\varepsilon_0 \approx$ 0.36V/nm) and $|E|$=200mV/nm ($|D|/\varepsilon_0 \approx$ 0.8V/nm), respectively. Because of ignoring screening effects in TDBG, the calculated regime of displacement field to produce isolated flat band would be relatively lower than experimental results. In our calculations, energies are shifted such that CNP is located at zero energy.

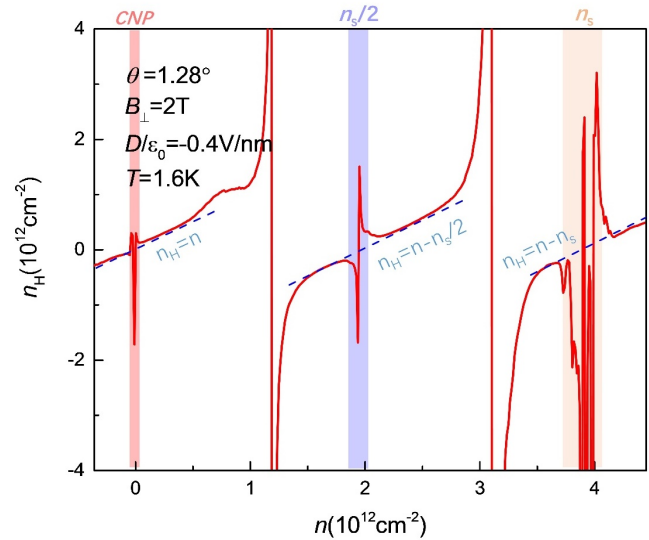

**Extended Data Fig. 4| Hall carrier density measurements in the 1.28°-device.** The Hall carrier density $n_H$=-1/($eR_H$) in the 1.28°-device is plotted as a function of gate-induced charge density. The data are acquired at magnetic field $B_\perp$=2T and $D/\varepsilon_0$=-0.4V/nm. Vertical colored bars correspond to various fillings of flat conduction band. The Hall carrier density switches types of Hall carriers at $n$=0, $n$=$n_s$/2 and $n$=$n_s$, and strictly follows $n_H$=$n$, $n_H$=$n$-$n_s$/2 and $n_H$=$n$-$n_s$ in the vicinity of correspondingly empty, half and full filling of flat conduction band. This behavior serves a definitely evidence of fully opened gap at half filling.

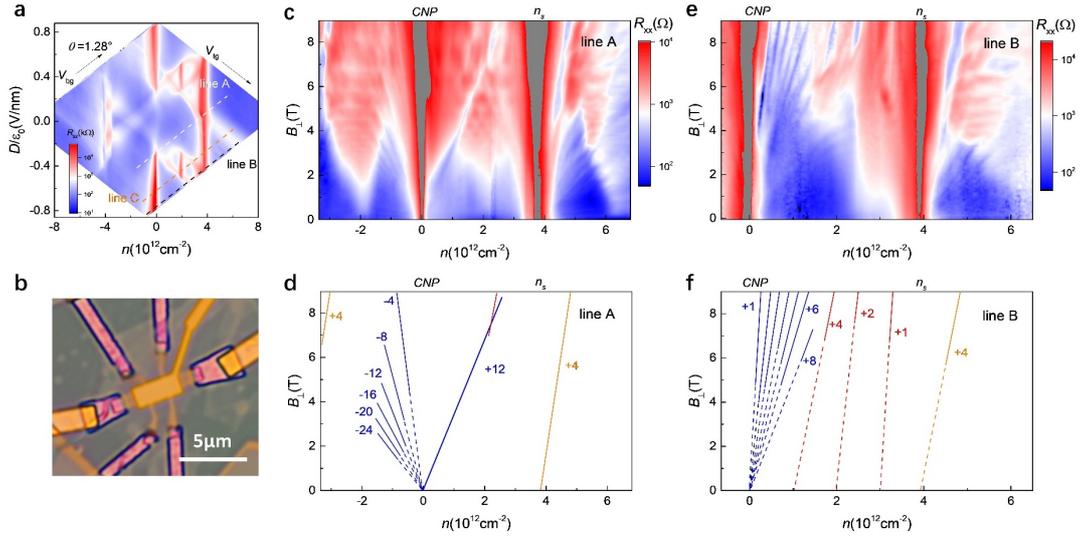

**Extended Data Fig. 5| Quantum oscillations in the absence of half-filling correlated state in the 1.28°-device. a**, Transformed resistance mapping plot from Fig. 1d as a function of $n$ and $D$. Line A, B and C denote the corresponding gate sweeping traces. Line C is the sweeping trace for Fig. 2b. **b**, Optical microscope picture of the 1.28°-device. **c**, **e**, Magneto resistance vs. carrier density $n$ in perpendicular magnetic field $B_\perp$ with gate voltage swept along trace A and B, respectively. At the same density, displacement field in **e** is always stronger than **c**. **d**, **f**, Schematic diagram of Landau levels observed in **c** and **e**, respectively. Landau levels originated from CNP and superlattice band edge are plotted with blue and orange colors, respectively. Fractal Landau levels are plotted with red colors.

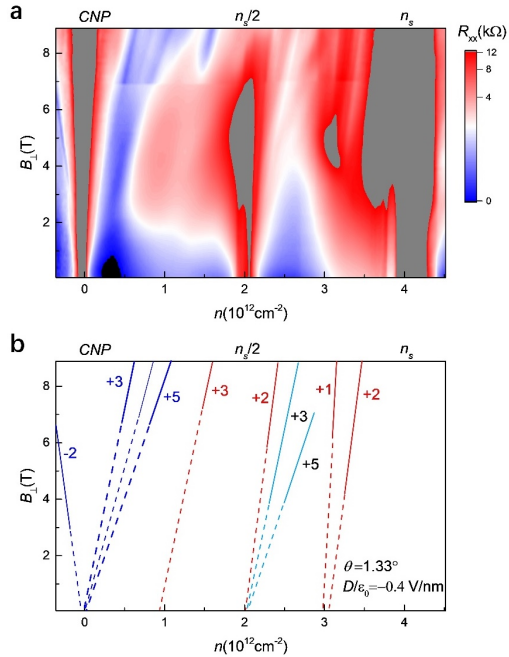

**Extended Data Fig. 6| Quantum oscillations in the 1.33°-device. a**, Magnetoresistance oscillations in perpendicular magnetic field $B_\perp$ varied from 0T to 9T and at $D/\varepsilon_0$=-0.4V/nm. **b**, Schematic of Landau levels observed in **a**. Dark blue lines, light blue lines and red lines track Landau levels fanning from CNP, Landau levels fanning from half filling and fractal Landau levels, respectively.

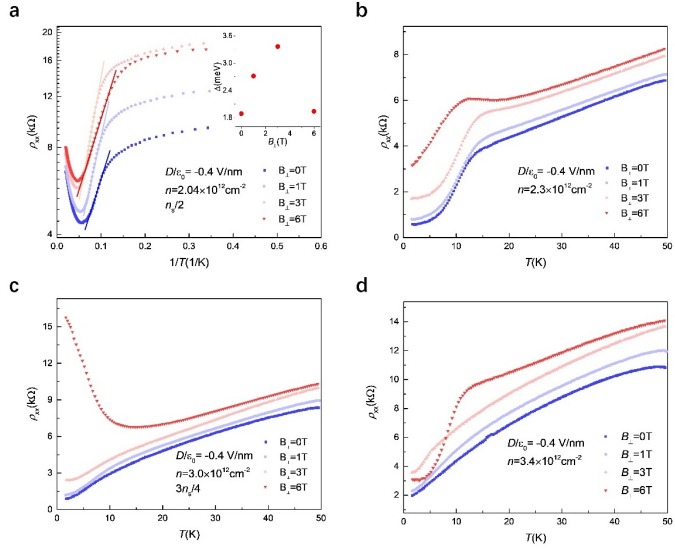

**Extended Data Fig. 7| Resistivity as a function of temperature at and near 1/2 and 3/4 fillings in various perpendicular magnetic fields. a, c,** Temperature-dependent resistivity behaviors rightly at half filling (**a**) and 3/4 filling (**c**). The inset figure in **a** shows thermal activation gaps at $B_\perp$=0T, 1T, 3T and 6T. The fitting is denoted by lines in the main figure according to Arrhenius formula. The 3/4-filling insulating state is induced at $B_\perp$=6T. **b, d,** $T$-dependent resistivity behaviors at electron-doped half filling (**b**) and 3/4 filling (**d**). All the data are acquired in the 1.33°-device.

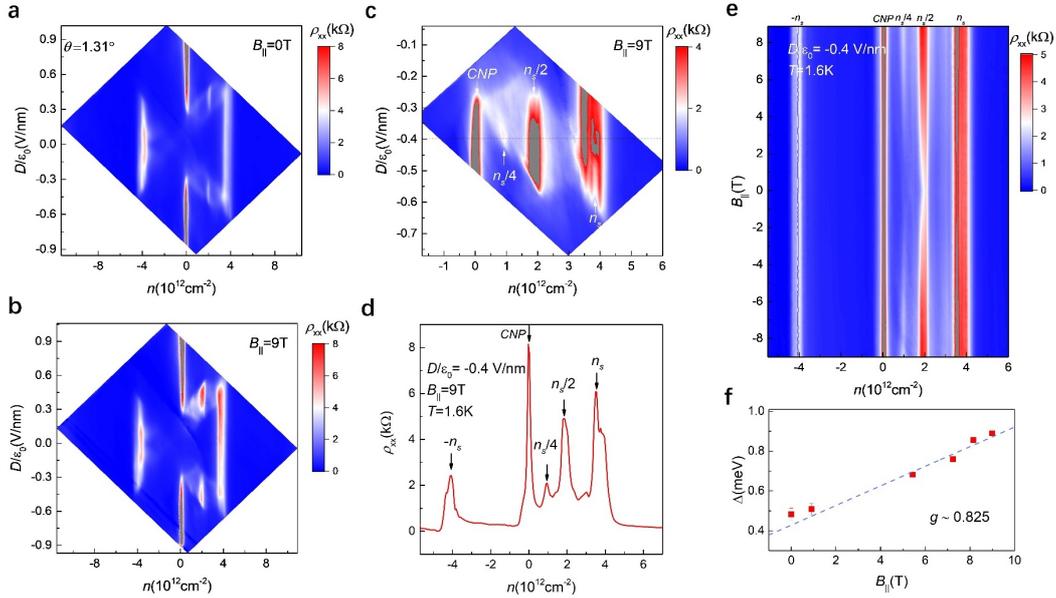

**Extended Data Fig. 8| Correlated insulators enhanced by parallel magnetic fields in the 1.31°-device. a, b,** Resistivity as a function of $D$ and $n$ at $B_\parallel$=0T(**a**) and $B_\parallel$=9T(**b**). **c,** Zoomed-in image for clear displaying of 1/4 and 1/2 filling insulators at $B_\parallel$=9T. **d,** Resistivity versus density $n$ at $D/\varepsilon_0$=-0.4V/nm corresponding to the dash line in **c**. **e,** $B_\parallel$ dependence of all insulating states, including $B_\parallel$-induced 1/4-filling and enhanced half-filling correlated states. **f,** A fitted effective $g$ factor according to the spin-Zeeman effect. The data in **f** show thermal-activation gaps at various $B_\parallel$. Error bars in **f** are estimated from the uncertainty in the range of simply activated regime.

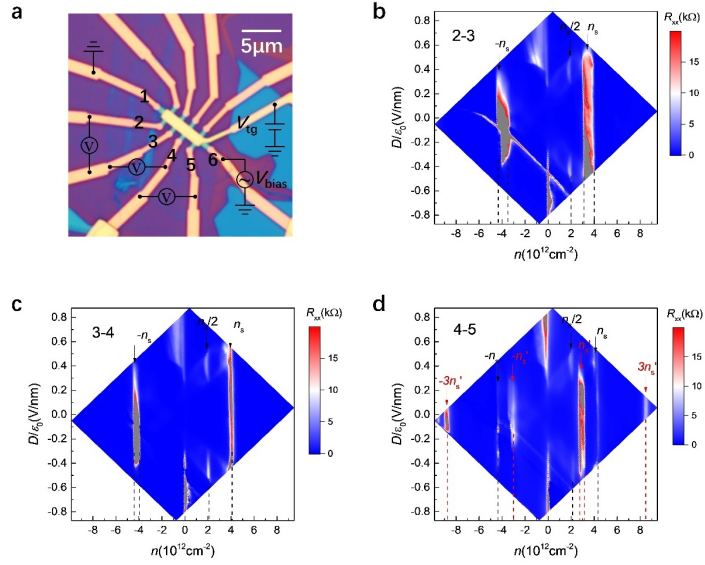

**Extended Data Fig. 9| Twist angle inhomogeneity in the 1.33°-device. a**, Schematic of measurement configuration and optical image of the 1.33°-device. **b, c, d,** Resistance color plot versus carrier density $n$ and displacement field $D$ at 1.6K acquired between contacts shown in **a**. We could extract twist angle $\theta=1.33°\pm0.01°$ as well as $\theta=1.11°\pm0.04°$ between contacts 4 and 5 according to the carrier density $n_s/2$ or $n_s$ in **d**. The errors here are estimated from the uncertainty in determining resistance peaks in **d**. The discussed transport data in main text and Methods for 1.33°-device were acquired between contacts 3 and 4.

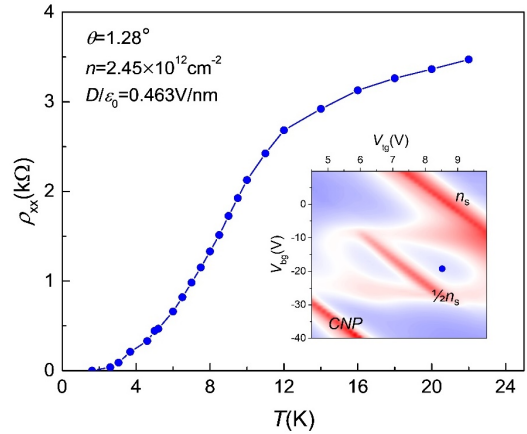

**Extended Data Fig. 10| Zero resistivity in the 1.28°-device.** Temperature-varied resistivity data were acquired at $n=2.45\times10^{12}\mathrm{cm}^{-2}$ and $D/\varepsilon_0=0.463\mathrm{V/nm}$, where a blue dot is located in the inset figure. The inset figure shows resistance mapping as a function of $V_{bg}$ and $V_{tg}$ for the 1.28°-device at $T=3\mathrm{K}$.